# Simultaneous column-based deep learning progression analysis of atrophy associated with AMD in longitudinal OCT studies

Adi Szeskin, PhD[1,2], Roei Yehuda, MSc[1], Or Shmueli, MD[3], Jaime Levy, MD[3], Leo Joskowicz, PhD[1],

[1]School of Computer Science and Engineering, The Hebrew University of Jerusalem, Jerusalem, Israel

[2]The Alexander Grass Center for Bioengineering, The Hebrew University of Jerusalem, Israel

[3]Department of Ophthalmology, Hadassah Medical Center, Jerusalem, Israel

**Word count**: 6,270

**Funding source**: Partial funding was provided by Integra Holdings Ltd.

**Disclosure**: None of the authors has any financial/conflicting interests to disclose.

**Correspondence**:

Dr. Adi Szeskin,

School of Computer Science and Engineering,

The Hebrew University of Jerusalem,

Jerusalem, Israel

E-mail address: adi.szeskin@mail.huji.ac.il

**Abstract**

**Purpose:** Disease progression of retinal atrophy associated with AMD requires the accurate quantification of the retinal atrophy changes on longitudinal OCT studies. It is based on finding, comparing, and delineating subtle atrophy changes on consecutive pairs (prior and current) of unregistered OCT scans.

**Methods:** We present a fully automatic end-to-end pipeline for the simultaneous detection and quantification of time-related atrophy changes associated with dry AMD in pairs of OCT scans of a patient. It uses a novel simultaneous multi-channel column-based deep learning model trained on registered pairs of OCT scans that concurrently detects and segments retinal atrophy segments in consecutive OCT scans by classifying light scattering patterns in matched pairs of vertical pixel-wide columns (A-scans) in registered prior and current OCT slices (B-scans).**Results:**

**Results**: Experimental results on 4,040 OCT slices with 5.2M columns from 40 scans pairs of 18 patients (66% training/validation, 33% testing) with 24.13±14.0 months apart in which Complete RPE and Outer Retinal Atrophy (cRORA) was identified in 1,998 OCT slices (735 atrophy lesions from 3,732 segments, 0.45M columns) yield a mean atrophy segments detection precision, recall of 0.90±0.09, 0.95±0.06 and 0.74±0.18, 0.94±0.12 for atrophy lesions with AUC=0.897, all above observer variability. Simultaneous classification outperforms standalone classification precision and recall by 30±62% and 27±0% for atrophy segments and lesions.

**Conclusions:** simultaneous column-based detection and quantification of retinal atrophy changes associated with AMD is accurate and outperforms standalone classification methods.

**Translational relevance**: an automatic and efficient way to detect and quantify retinal atrophy changes associated with AMD.

## 1. Introduction

Disease progression of retinal atrophy associated with age-related macular degeneration (AMD) requires the objective and accurate quantification of the retinal atrophy changes on clinical images. AMD affects an estimated 196 million people worldwide [1,2]. It is the leading cause of blindness in people older than 65 years in the Western world, affecting 1 in 4 adults over the age of 75. Here within, we refer to as macular atrophy, or more simply as atrophy.

Accurate quantification of AMD progression is a key unmet clinical need. As macular atrophy progresses, there may be profound effects on visual function. Quantifying AMD progression is essential to objectively evaluate treatment efficacy, to establish a prognosis, and to support new drugs and treatments development. While there is currently no treatment to repair damaged retinal pigment epithelium (RPE) and photoreceptor cells in eyes with advanced atrophic AMD, numerous efforts are underway to develop treatments to slow or stop AMD progression, e.g., stem cell therapy, intravitreal complement inhibitor injection, and electrophysiology [3].

AMD is classified as early, intermediate, and late AMD based on the macula appearance. Advanced AMD is either wet (neovascular) or dry (non-neovascular) – about 8 out of 10 people with AMD have the dry form [3]. Dry AMD is further classified into four categories of increasing severity: iORA, cORA, iRORA (non-complete macular atrophy) and cRORA, complete RPE and outer retinal atrophy. cRORA is separately quantified as the endpoint of advanced dry AMD [4]. These atrophies have subtle differences, making their classification difficult even for experienced ophthalmologists.

Volumetric spectral domain optical coherence tomography (SD-OCT, here within OCT) has been established as the most accurate imaging method for early diagnosis and reliable follow-up of AMD progression and its stages [5]. However, OCT scans consist of up to hundreds of slices (B-Scans) and their analysis has not been standardized. Their examination is tedious, time-consuming, requires expertise and is subject to significant observer variability. This is exacerbated in follow-up studies, where atrophy changes are identified by comparing OCT scans acquired at different times. Thus, manual delineation and quantification of atrophy and atrophy changes in OCT slices is not routinely performed in the clinic.

Follow-up disease progression evaluation of dry AMD on longitudinal OCT scans is significantly more challenging than evaluation on individual OCT scans: it requires finding, comparing, delineating and quantifying possibly subtle atrophy changes on two or more consecutive pairs (prior and current) of unregistered OCT scans acquired at different times (Fig. 1). While longitudinal atrophy evaluation can be obtained by automatic standalone evaluation of each OCT scan in the series and comparison of the results [6,7], standalone analysis has three drawbacks: 1) the reported atrophy changes are not spatially localized since the OCT scans are not registered; 2) the atrophy changes detection requires atrophy classification in each OCT scan: it is not determined by the relative change of the atrophy appearance in both OCT scans which may be subtle but significant; 3) the extent and boundaries of the atrophy changes may be fuzzy and may only be determined by comparative analysis. These problems worsen for patient studies with multiple OCT scans with different resolutions and from different scanners.

This paper presents a novel, fully automatic end-to-end pipeline for radiomics-based disease progression analysis of consecutive OCT studies of patients with advanced dry AMD. The key novel premise of our method is that the **simultaneous** classification of light scattering patterns in **spatially matched pairs** of vertical pixel-wide **columns** (A-scans) in registered prior and current OCT slices (B-scans) where atrophy appears yields superior recall and precision than single, standalone atrophy detection and segmentation in each OCT scan followed by atrophy changes detection and analysis. This is because simultaneous atrophy detection identifies the differences in the outer retinal degeneration and the light transmission in the choroid layer beneath the retina of **pairs of columns** in each OCT slice in the **same spatial location**.

2. **Previous work**

Various methods have been proposed for the evaluation, classification, and segmentation of wet and dry AMD atrophy in individual OCT scans. Model-based approaches include graph-based and dynamic programming techniques for segmenting the horizontal retinal layers [8,9,10]. Machine and deep learning methods have been recently proposed for pathologies and structures in individual OCT scans. Kurmann et al [11] describes a CNN-based method for identifying and classifying retinal biomarkers in OCT scans. Lee et al [12] describes a method for binary healthy/atrophied classification of individual OCT slices with a modified VGG16 CNN trained on 48,000 manually annotated OCT slices. Ji et al [13] presents a deep learning binary classification method for macular atrophy segmentation based on sparse one-pixel OCT columns autoencoders. Venhuizen et al [14], De Fauw et al [15] and Fang et al [16] describe U-Net based methods for retina thickness and retinal tissue quantification. Shah et al [17] presents a CNN-based method for the segmentation of retinal layers in OCT slices. Dong et al [18] describes a multi-path block structure for GA segmentation. Derradji et al [19] describes a CNN method for RPE and ORA detection and segmentation. Szeskin et al [7] describes a column-based method for the standalone atrophy segment and atrophy lesion detection and segmentation. None of these methods uses information from previous patient OCT scans.

A few papers address AMD disease progression based on longitudinal OCT studies. Bogunović et al [20] presents a hybrid model-based and machine learning method for retinal disease progression estimation in longitudinal OCT scans. Banerjee et al [21] describes a method for AMD progression prediction that combines Recurrent Neural Network and Random Forest classification. Both methods address atrophy type classification, not atrophy segmentation and quantitative atrophy changes analysis. Schmidt-Erfurth et al [22] presents a method for predicting early AMD disease conversion based on individual retinal layer segmentation. None of these methods performs atrophy segmentation and OCT scan registration.

More broadly, various methods have been proposed for the simultaneous identification and segmentation of structures and lesions in longitudinal CT and MRI in the liver, lungs and brain. Birenbaum et al [23] describes a method for sclerosis lesion detection and segmentation in pairs of MRI scans that independently finds lesions in each scan with a standalone CNN and merges them. Vandewinckele et al [24] presents a method for the concurrent segmentation of head/neck structures in longitudinal CT scans. It combines features from registered CT scans intensities using cross-sectional CNN.

In summary, while various methods have been proposed for the identification, classification, disease progression and prognosis analysis of AMD in OCT scans, none of them addresses the simultaneous identification, quantification and tracking of cRORA segments and atrophy lesions changes over time on registered OCT scans.

## 3. Method

The key novelty of our new method is the simultaneous pairwise analysis of atrophy lesion and atrophy lesion changes in registered, consecutive OCT studies. The analysis is performed with a novel multi-channel column-based CNN network trained on registered pairs of OCT scan columns. The network leverages on the additional spatial and voxel intensity information (grey pixel values of two OCT columns instead of one) in pairs of registered matching 3D OCT column patches in the prior and current OCT scans. The classification is based on both the relative healthy and atrophy column appearance differences and on their individual differences.

The simultaneous pairwise analysis is based on the classification of columns in each OCT slice showing dry AMD atrophy described in our previous work [7]. The AMD classification of each column is performed on a column patch formed by adjacent neighboring columns to the left and right of the central column. Contiguous atrophy columns form atrophy segments – there may be more than one atrophy segment per OCT slice. Atrophy segments are projected onto a binary atrophy segments matrix that is used to identify and segment atrophy lesions in the IR image and to perform measurements (Fig. 1). The advantage of the column-based method over the existing retinal layer and retinal patch methods is that the columns method uses all the retinal layers and the scattering pattern below the RPE layer. It obviates the need for retinal layer segmentation and simplifies the annotation of atrophy lesion segments to 1D intervals.

Fig. 2 shows a novel fully automatic end-to-end pipeline for the simultaneous detection and quantification of atrophy and atrophy changes associated with AMD in consecutive pairs of OCT scans of a patient. The inputs are the prior and current OCT studies (IR image and OCT scan); the outputs are the atrophy segments in both OCT scans, the atrophy lesions in the IR images, and an atrophy changes report.

**3.1 OCT columns registration**

Pairwise analysis of the prior and current OCT scans and their corresponding OCT columns requires establishing a common reference frame to account for the different patient head positions. The OCT columns are matched using the known spatial relation between the OCT scan and the IR image provided by the OCT scanner. The matching is computed by landmark-based 2D rigid registration of the prior and current IR images. The landmarks are points on the retina blood vessels in both images. Retinal blood vessels are known to be accurately identifiable and stable over time, so they are reliable registration features [25]. The registration accounts for the resolution and field of view (FOV) differences in the OCT scans and IR images. While some blood vessels may be occluded by atrophy in the IR images in the center of the retina, enough blood vessels are always visible in the periphery of the retina.

The OCT columns registration is performed on **matching** prior and current OCT slices. It consists of three transformations (Fig. 3): 1) prior OCT scan columns to prior IR image; 2) prior IR image to current IR image; 3) current IR image to current OCT scan columns. The first and third transformations are known; the second has to be computed.

Formally, let $(x, y, z)$ be a coordinate system in which the $xy$ plane is the IR image plane and the $yz$ plane is an OCT slice. The $x$ coordinate is the OCT slice number, the $y$ coordinate is the OCT column location on the OCT slice, and the $z$ coordinate is the location on the OCT column. Let $(x,y)^P_{OCT} = (x^P_{OCT}, y^P_{OCT})$ and $(x,y)^C_{OCT} = (x^C_{OCT}, y^C_{OCT})$ be the coordinates of an OCT column in the prior ($P$) and current ($C$) OCT scans in the IR image plane. Let $T^P_{OCT\ to\ IR}: (x,y)^P_{OCT} \to (x,y)^P_{IR}$ and $T^C_{OCT\ to\ IR}: (x,y)^C_{OCT} \to (x,y)^C_{IR}$ be the transformations that map the prior and current OCT scan to the IR image, and let $T^{P\ to\ C}_{OCT}: (x,y)^P_{OCT} \to (x,y)^C_{OCT}$ be the 2D rigid transformation between the prior and the current IR images:

$$T^{P\ to\ C}_{OCT} = {T^C_{OCT\ to\ IR}}^{-1} \circ T^{P\ to\ C}_{IR} \circ T^P_{OCT\ to\ IR}$$

Note that the transformations are invertible, i.e., the OCT scan columns of the prior and current OCT scans can interchangeably be defined as the registration basis.

The transformation $T^P_{OCT\ to\ IR}$ is computed from the location of the OCT column to the OCT scans' resolutions in both modalities. Let $\hat{x}^P_{OCT} = \frac{x^P_{OCT}}{n^P}$, $\hat{y}^P_{OCT} = \frac{y^P_{OCT}}{w^P_{OCT}}$ be the normalized OCT column slice number and the column location on the OCT slice, where $n^P$ is the number of slices and $w_{OCT}^P$ is the OCT slice width in the prior OCT scan. Then $T_{OCT\ to\ IR}^P((x,y)_{OCT}^P) = (h_0 + h_{IR}^P \cdot \hat{x}_{OCT}^P, w_0 + w_{IR}^P \cdot \hat{y}_{OCT}^P)$ where $h_{IR}^P$ and $w_{IR}^P$ are the height and width of the prior IR image and $(h_0, w_0)$ is the location of the bottom left corner of the OCT scan FOV in the IR image. $T_{OCT\ to\ IR}^C$ is defined and computed similarly.

The transformation $T_{OCT}^{P\ to\ C}$ is computed by landmark-based 2D rigid registration between the prior and the current IR images in three steps: 1) detection of retinal blood vessels based on the IR image gradient by Frangi filtering for continuous edge detection [26]; 2) automatic identification of landmark points in the retinal blood vessels image by ORB feature detection [27]; 3) matching landmark points in both IR images using RANSAC with a 2D rigid transformation that minimizes the sum of the squared distances between the paired landmark points. Points whose nearest neighbor in the prior IR are at a pre-defined distance $\leq d_{reg}$ are discarded as outliers.

The registration accuracy is automatically evaluated by comparing the difference of the blood vessels segmentation before and after the transformation. When the registration is deemed inaccurate (accuracy above a predefined threshold), the transformation is computed from at least three landmark points that are manually selected in each IR image with ***OCT-SIM***. Since the OCT slices do no cover the entire field of view, the matched OCT columns are the nearest columns in the matching prior and current slices and not the exactly same columns, which would require undesirable column pixels grey value interpolation.

### 3.2 Simultaneous atrophy columns detection

The atrophy columns detection in the prior and current OCT scans is performed with a novel simultaneous multi-channel column-based CNN network architecture (Fig. 4). The inputs are two registered prior and current 3D OCT column patches, and optionally, a prior binary atrophy segments matrix (prior mask). The prior mask can be obtained automatically as described in [7] or manually defined by the user. The output is a pair of bits indicating the absence or presence of dry AMD atrophy in the OCT columns.

The 3D column patches pairs are constructed in two steps: 1) individual column denoising of the prior and current OCT scans; 2) 3D column patch pair generation for each OCT slice. Denoising is necessary since the OCT scanning process produces both vertical and lateral light scattering that affects neighboring pixels. It consists of local average smoothing by convolution with a normalized $5 \times 5$ box filter [7]. For each matched prior and current OCT slice of height $h_{OCT}$ and width $w_{OCT}$, $m = \left\lfloor \frac{w_{OCT}}{s} \right\rfloor$ two 3D column patches of size ($h_{OCT} \times w \times 3$) where $h_{OCT}$ is the height of the OCT scan, $w$ is the width of the column, and 3 are the previous, current, and next OCT slices. are generated, where $s < w$ is the overlap stride between the column patches (columns at the OCT scan edges that do not fit into a column patch are ignored).

The simultaneous column-based CNN network N3 consists of two identical standalone column feature extraction (FE) networks N1 (PRIOR and CURRENT) and a pairwise column FE network N2 (PAIR). The FE networks perform individual and pair features extraction of the individual prior and current column patches, which are then processed by fully connected (FC) layers whose outputs are vectors of size $32 \times 1$. Optionally, the prior mask is processed by an FC layer that outputs a vector of size $8 \times 1$. The prior column bit is computed with single-channel FC layer from the prior features vector; the current column bit is computed with a flattened four-channel FC layer from the prior, current, pair, and prior mask feature vectors. The bits are then combined into a pair.

The standalone column FE network N1 is as described in [7]. It consists of two sequences of a convolutional layer and a max-pooling layer followed by an FC and a drop-out layer. The input layer ($h_{OCT} \times w \times 3$) and the threshold of the last softmax layer *th* is determined by analyzing the ROC and the Precision-Recall curves of the pairs validation dataset. Inference is performed on all pairs of matched prior and current patches. The output is the OCT column classification in each of the prior and current OCT slices. FC layers have *RELU* activation; the last FC has *softmax* activation; it outputs a features vector of size $256 \times 1$. The kernel sizes are: first convolutional layer $496 \times 7 \times 32$, convolution $3 \times 2$; first pooling layer $165 \times 3 \times 32$, max-pooling $3 \times 2$; second convolutional layer $165 \times 3 \times 64$, convolution $3 \times 1$; second pooling layer $55 \times 3 \times 64$, max-pooling $3 \times 1$; FC layer $256 \times 1$, drop-out layer $256 \times 1$, drop-out rate 0.4. The pairwise column FE network N2 is identical to N1 except for its first convolution layer, which inputs $2 \times (496 \times 7 \times 32)$ column patches pairs.

The network training proceeds as follows. First, a standalone column FE network N1 together with the features FC layer is trained with an annotated individual OCT scan training dataset D1 as described in [7]. Since there are many more columns without atrophy than with atrophy, only a subset of non-atrophy 3D column patches are selected to balance the data. Augmentation is performed on all 3D column patches with atrophy by horizontal mirroring and small in-plane rotations of $\pm 2º$. Next, the pairwise column FE network N2 is trained with an annotated pairs OCT scan training dataset D2. Finally, two simultaneous column networks N3 and N3 with prior mask, are trained by keeping fixed (freezing) the weights of the FE networks N1 and N2 and training the three (four with prior mask) flattened FC layers on the dataset D2.

For inference, no 3D column patch augmentations are performed. The output of the prior and current FC layers are continuous values in the range of [0,1] on which binary thresholding is performed with a pre-set threshold *th* that is empirically determined by analyzing the ROC and the Precision-Recall curves of the pairs validation dataset. Inference is performed on all pairs of matched prior and current patches. The output is the OCT column classification in each of the prior and current OCT slices.

### 3.3 Atrophy segments and lesions computation

The computation of the atrophy lesion segmentation, the extraction of the IR image pixel size (in $\mu m$), and the location of the fovea, are performed as in [7]. Briefly, atrophy lesion segmentations are computed by projecting the OCT segments matrix onto the IR image using the known transformations $T^P_{OCT\ to\ IR}$ and $T^C_{OCT\ to\ IR}$. The result is two 2D binary OCT atrophy segments matrices of size $n^P \times w^P_{OCT}$ and $n^C \times w^C_{OCT}$, derived directly from the atrophy segments in each OCT slice, where $n^P, n^C, w^P_{OCT}, w^C_{OCT}$ are the number of OCT slices and their respective widths in the OCT scans.

The atrophy lesions are then identified and segmented in the prior and current IR by computing the OCT scan FOV axis-aligned rectangle in the IR image, projecting the atrophy segments in the OCT slices onto the IR image, and identifying and segmenting the atrophy lesions. Each connected component corresponds to an atrophy lesion whose segmentation is the union of all pixels in the component. The common FOV is the intersection of the prior and current FOVs.

### 3.4 Atrophy and atrophy changes measurement

The atrophy and atrophy changes report includes the clinically relevant measurements of the lesion atrophy extent in both the prior and current scans as defined in [28]. It lists the total lesions area A for all lesions whose area is >0.05 mm$^2$ and the lesion area in concentric disks centered at the fovea of diameters of 1,3,6mm. Six additional parameters are computed: 1) lesion (cumulative) perimeter P defined as the length of the (cumulative) circumference; 2) lesion (cumulative) circularity, C = $4\pi A/P^2$ ; 3) focality index, the number of lesion atrophy connectivity components whose area is >0.05 mm$^2$; 4-5) Feret-max and Feret-min caliper defined by the maximal (minimal) perpendicular distance between parallel tangents on opposite lesion sides; 6) minimal lesion distance from the fovea.

Disease progression is quantified with areal and directional atrophy rates. Areal progression rates [mm$^2$/year] are computed for the entire retina and for concentric disks centered at the fovea of diameters of 1,3,6 mm divided into quarters. Directional progression rates [mm/year] are computed for nine directions: superior, inferior, nasal, temporal directions, intermediate directions, and inner radial direction i.e., towards the fovea.

**3.5. Longitudinal atrophy changes analysis in OCT studies**

The simultaneous atrophy changes analysis pipeline is used for the detection and quantification of cRORA and atrophy changes in longitudinal OCT studies as follows.

Let $S = <S_1, \ldots, S_n>$ be a sequence of consecutive OCT studies ($n > 2$), where $S_n$ is the current, most recent OCT study and $S_1$ is the first, oldest OCT study. The sequence $S$ is processed in a cascaded manner, starting with the oldest OCT pair, $(S_1, S_2)$, processed with the simultaneous column-based model $M3$ to produce their corresponding atrophy lesion segmentations $(LS_1, LS_2)$. Subsequent OCT pairs $(S_i, S_{i+1})$ are processed with the model $M3$ with prior mask, where the prior mask $LS_i$ is the computed lesions of OCT study $S_i$, until the final prior and current OCT pair $(S_{n-1}, S_n)$ is reached. When validated atrophy lesion segmentation priors are available, they are used instead of the computed ones. At the end of this process, the consecutive pairwise atrophy lesion and lesion changes are directly aggregated into a longitudinal analysis.

**4. Experimental results**

To evaluate our methods, we implemented them, collected clinical ophthalmic studies, annotated their OCT scans, and conducted four experimental studies. We describe each next.

**4.1 Datasets**

The ophthalmic studies were retrospectively obtained from the Hadassah University Medical Center (Ein Kerem, Jerusalem, Israel) by the two co-author ophthalmologists during the routine clinical examination of patients. The studies were selected from patients that were diagnosed as having macular atrophy. The studies were acquired with a SPECTRALIS$^{TM}$ OCT scanner (Heidelberg Engineering, Germany). The IR image sizes are 496×496, 20×20 $\mu m^2$. The OCT scans consist of 25-80 slices of sizes 496×1024 and 496×1536 pixels per slice and resolutions of 4×6 $\mu m^2$.

We created two studies datasets $D1$ and $D2$ for cRORA. Dataset $D1$ consists of 106 individual OCT studies (OCT scan and IR image) from 18 patients with macular degeneration. Dataset $D2$ consists of 40 pairs of prior and current OCT studies from 18 patients with macular degeneration. The mean time elapsed between prior and current OCT studies of a patient is 24.13±14.0 months (range 3-59).

### 4.2 Manual Annotations

For $D1$, cRORA annotation in 106 cases was performed on the OCT slices as in [7]. In total, 5,207 OCT slices were examined. Atrophy was identified in 2,952 OCT slices, in which 5,111 atrophy segments corresponding to 1,046 atrophy lesions were detected. On average there were 48.3±37.5 atrophy segments (range 0–294) per OCT scan, corresponding to an average of 9.9±8.2 atrophy lesions (range each with 0–39). This resulted in an annotated dataset of 689,256 (10%) cRORA columns out of the total 6,815,232 OCT columns in all scans. Each OCT slice had a mean of 134.6±141.1 atrophy columns (range 3-886).

For $D2$, cRORA annotation in all 40 pairs was performed manually on the OCT slices with *OCT-SIM* by viewing prior and current OCT slices simultaneously (Fig. 1). All pairs of OCT scans were initially annotated by the first ophthalmologist co-author (OS). The second senior ophthalmologist co-author (JL) then independently annotated 9 pairs and validated and corrected when needed the segmentations of the first ophthalmologist for all other 31 pairs. In total, 4,040 OCT slices of the 80 OCT scans were examined. Atrophy was identified in 1,998 OCT slices, in which 3,732 atrophy segments corresponding to 735 atrophy lesions were detected. On average, there were 46.6±27.3 atrophy segments (range 0–106) per OCT scan, corresponding to an average of 9.2±8.1 atrophy lesions (range 0–29) per OCT scan. The mean±std areal change factor between the prior and the current OCT studies is 2.1±1.5 (range 1.1-9.1). This yielded a dataset of 450,430 (8.5%) cRORA columns out of 5,284,864 OCT columns in all scans. Each OCT slice had an average of 120.7±124.7 atrophy columns (range 3-783).

## 4.3 Evaluation Metrics

We quantify the performance of our method with respect to the manual ground-truth annotations with the following metrics. For atrophy segment and atrophy lesion detection, we use Precision and Recall. For atrophy segment and atrophy lesion segmentation, we use the $F_1$ score [29], the Average Symmetric Surface Distance (ASSD) and the Symmetric Hausdorff Distance (SHD). We also compute the receiver operating characteristic curve (ROC), the AUC (area under the ROC), the Precision-Recall curves, and Confusion Matrices.

## 4.4 Models

We trained seven models as follows. Model $M1$ is the standalone column FE network $N1$ followed by a $256 \times 1$ dropout layer and a one-bit FC layer trained on individual OCT scans dataset $D1$ [7]. Models $M3\_M2$ and $M3\_M2\_P$ (without and with prior mask) are the simultaneous columns detection network $N3$ without the pairwise column FE network N2 and its subsequent pair features flattened FC layer trained on OCT scan pairs dataset $D2$ by keeping fixed (freezing) the weights of the FE networks N1. Models $M3\_M1$ and $M3\_M1\_P$ (without and with prior mask) are the simultaneous columns detection network $N3$ without the prior and current standalone column FE networks N1 and its subsequent prior and current features FC layer trained on OCT scan pairs dataset $D2$ by freezing the weights of the FE networks N2. Models $M3$ and $M3\_P$ (without and with prior mask) are the simultaneous columns detection network $N3$ trained on OCT pairs dataset $D2$ by freezing the weights of the FE networks $N1$ and $N2$.

The standalone model $M1$ was trained and cross-validated on 5,207 individual OCT slices with a total of 6,815,232 columns from 106 OCT studies of 18 patients, all from $D1$. All the other simultaneous models were trained and cross-validated on 2,618 pairs of OCT slices with a total of 2,131,048 columns from 34 pairs of OCT studies from 16 patients from $D2$. The test set for all models was 420 pairs of OCT slices with a total of 342,016 columns from 6 pairs of OCT studies from 4 patients from $D2$ (for $M1$, the pairs were considered individually). The patients in the test set were chosen randomly so that their mean areal change factor (from prior to current) is closest to the mean for all pairs so it will still reflect the dataset characteristics.

The models were trained with $Adam$ optimizer, learning rate of 0.001, a batch size of 100 column patches, and the $F_1$ loss function (see [7] for comparison with other losses). Training was performed with 4-fold cross validation for 1,000 epochs. In each fold, the weights were chosen by the minimal mean $F_1$ loss on its validation dataset. The fold which yields the $F_1$ score closest to the mean $F_1$ on the validation score for all folds excluding folds with $F_1$ scores that deviate from the mean by more than one std were chosen.

### 4.5 Experimental Studies

We conducted four experimental studies as follows.

**Study 1: Manual cRORA segmentation in OCT slices.** The inter-observer annotation variability was quantified by independent manual annotation by the two coauthor ophthalmologists of 834 OCT slices from 9 OCT study pairs in $D2$ where 4,096 atrophy segments with a total of 83,043 atrophy columns from 186 atrophy lesions were detected.

The mean difference between the two clinician annotations was +1,628±1,876 columns (range −1,442 to +4,994), which yields a mean difference in atrophy burden of +4.3±4.9% and a mean $F_1$ score of 0.71±0.10 (range 0.46–0.80). With the senior ophthalmologist as reference, the mean ASSD and SHD were 0.23±0.41 mm and 1.79±2.19 mm. The mean precision and recall of atrophy segments detection were 0.76±0.18 and 0.88±0.14 respectively. The mean precision and recall of atrophy lesion detection were 0.68±0.30 and 0.89±0.13 respectively. The relatively low $F_1$ score and precision were attributed to the fuzzy cRORA boundaries, which yield differences in their interpretation and annotation.

We quantify the inter-observer variability of the atrophy and atrophy changes measurements. These measurements were automatically computed from the manual cRORA annotations independently performed by the ophthalmologists on the same 9 OCT studies pairs. Of the 28 measurements, we report the five most significant ones that quantify the atrophy change. For each measurement, we compute the measurement difference RMSE (root mean squared error) and standard deviation (std). Table 1 (top) lists the results.

**Study 2: Automatic cRORA segmentation in OCT slices**. We compute cRORA segments on OCT slices with the standalone ($M1$) and the simultaneous column-based models ($M3\_M2$, $M3\_M1$, $M3$) with and without a prior mask.

Table 2 summarizes the results. Figure 5 show an example of the results. The model $M3$ yields the best results. For the atrophy extent, the mean±std difference between the manually reference ground-truth cRORA burden and the automatically derived one was +8.2±7.8% (range 0.7-21.4) with the prior mask, and +13.0±6.7% (range 4.0–22.6) without the prior mask. The mean $F_1$ score is 0.77±0.07 (range 0.66–0.83) with the prior mask and 0.74±0.08 (range 0.62–0.83) without the prior mask. The atrophy detection mean precision and recall for atrophy segments are 0.91±0.06 and 0.89±0.06 with the prior mask and 0.90±0.09 and 0.95±0.05 without the prior mask, and for atrophy lesions 0.84±0.17 and 0.83±0.25 with the prior mask and 0.74±0.18 and 0.94±0.12 without the prior mask. These results show that the automatic simultaneous column-based classification is as accurate and reliable as the manual segmentation, within the observer variability of Study 1.

ROC, Precision-Recall curves, and confusion matrices of the simultaneous models, and particularly $M3$, clearly outperform the standalone model $M1$: $M3$ has an AUC of 0.897 and a mean precision of 0.64, better that the other classifiers. For the confusion matrices, model $M3$ reaches the most balanced results. The results of $M3\_M2$ achieving a slightly higher true negative rate (TNR) of 83% vs. the 81% achieved by model $M3$ without the prior mask; its true positive rate (TPR) is 85% which is significantly lower than the 93% of model $M3$.

**Study 3: OCT column registration accuracy.** We quantify the accuracy of the OCT column registration (the first step of the pipeline) with three measures: 1) landmark-based 2D rigid registration of the prior and current IR images; 2) prior and current OCT slice selection matching; 3) prior and current OCT column selection matching within an OCT slice.

We randomly selected 20 pairs from dataset $D1$ and had an expert ophthalmologist manually select and pair 5-10 matching landmark points (fiducials) in the current and prior IR images. The ground truth 2D rigid transformations were computed from them with RANSAC. To compute the registration accuracy, new landmarks were automatically computed and selected on the prior image. Their corresponding points in the current image (if they exist) were automatically computed by projecting them onto the IR image, finding by RANSAC the matching points that are bot fiducial landmarks. The registration accuracy was the target landmarks RMSE.

The mean target registration error is 2.22±2.11 pixels, with 1.23±1.27 and 1.56±1.57 on the $x$ and $y$ axis. Since the IR image $xy$ resolution is 20×20 $\mu m^2$ and the $xz$ resolution of the OCT slices is 4×6 $\mu m^2$ with $z$ spacing of 140 $\mu m$, there is one OCT slice in every 7 (140/20) IR pixels. Thus, the accuracy of the OCT slice selection is 100% since the mean (max) $y$ error in the IR image is 1.3 (max=5.11) pixels, which is 7. The prior and current column selection accuracy is determined by the mean (maximum) $x$ error in the IR image: since there are 5 OCT columns per IR pixel (20/4) and the error is 2.22 (max=6.12), the mean (max) prior and current OCT column offset is 11.1(max=30.6), within the observer variability.

**Study 4: Clinical measurements.** This study quantifies the accuracy of the clinical measurements auto automatically computed from the cRORA lesions segmentations generated by the model $M3$ (without prior mask). For each measurement, the RMSE and std of the differences between the measurements are computed from: 1) the manually annotated reference ground-truth cRORA lesions segmentations of 834 OCT slices from 9 prior and current OCT studies pairs (Study 1), and; 2) the atrophy lesions segmentations for 40 prior and current OCT studies pairs computed with the model $M3$.

We selected the 5 out of 28 most clinically relevant measurements for the quantification of the atrophy progression between the current and prior OCT studies: 1) change in focality index; 2) atrophy area progression; 3-4) mean inward (towards the fovea) and outward (away from the fovea) directional progression rates computed from the nine directional progression rates in the nasal, temporal, superior, inferior direction; 5) mean Feret diameters progression rates.

Table 1 (bottom) summarizes the results. The measurements computed from the atrophy lesion segmentations generated by model $M3$ without prior mask classifier have RMSE and std values that are smaller than the inter-observer variability, with the only exception of the RMSE for area progression rate (0.90 vs. 0.74), whose value is acceptable.

## 5. Discussion

Our approach provides a fully automatic, end-to-end pipeline for the key tasks of simultaneous detection and quantification of time-related atrophy changes associated with AMD in pairs of consecutive prior and current OCT scans of a patient.

Our experimental results indicate that the simultaneous column-based model trained on pairs of pairs of prior and current OCT scans of a patient *outperforms* the standalone column-based model, achieving expert performance for cRORA detection and segmentation. For the simultaneous column-based model $M3$ without and with the prior mask, the mean $F_1$ score is 0.74±0.08 and 0.77±0.07, both better than the standalone model $M1$ mean $F_1$ score of 0.72±0.07 and of the observer variability, with a mean $F_1$ score of of 0.71±0.10. It significantly improves the mean precision of the standalone classifier by 30±62% and 27±0% for atrophy segments and atrophy lesions, respectively: 0.91±0.06 vs. 0.70±0.16 and 0.84±0.17 vs. 0.66±0.17.

The accuracy of the five most clinically relevant atrophy lesion progression measurements is also higher than the manual observer variability. The RMSE±std of the automatic vs. the manual measurements of the focality index is 1.29±0.82 vs. 1.76±1.75, of the area progression rate is 0.74±0.73 mm$^2$/year vs. 0.90±0.64 mm$^2$/year, of the mean outward and inward directional progression rates is 0.16±0.18 mm/year vs.

0.44±0.41 mm/year and 0.06±0.06 mm/year vs. 0.11±0.10 mm/year, and of the mean Feret diameters progression rate is 0.29±0.27 mm/year vs. 0.64±0.57 mm/year. On average, the mean automatic RMSE is better by 34±44% than the manual one. This shows that the automatic measurements are reliable.

Our method is unique in that it classifies retinal atrophy and its changes based on the light scattering pattern of matched pairs of OCT columns in the prior and current OCT scans: it includes all retinal layers instead of relying on their presence or absence and on the thickness of each individual retinal layer in each OCT scan. Also, our method is generic in that it only relies on manually labeled datasets used for training. Thus, it can be directly applied to dry AMD atrophy categories (iORA, cORA, IRORA) by obtaining ground truth labeling for each.

The advantage of our column-based method is that the manual labeling is performed on pairs of OCT atrophy segments, each requiring only a left and right column; each atrophy segment contains several hundreds of columns. This yields ~×100-1,000 more labeled data than the manual labeling of retinal layers on which existing deep learning methods rely. Thus, our method requires a small set of annotations to achieve expert-level performance: the simultaneous column-based models were trained on 34 prior and current OCT studies pairs in contrast to other deep learning methods based on retinal layers segmentation [11,30], which require ~×10-100 more annotated data which is more time-consuming to acquire.

The novelties of our method are: 1) a fully automatic, end-to-end pipeline for the simultaneous detection and quantification of time-related atrophy changes associated with AMD in pairs of consecutive OCT scans of a patient; 2) a novel simultaneous column-based CNN network architecture for pairwise classification of matched prior and current OCT columns; 3) accurate segmentation of complete retinal dry AMD atrophy in pairs of OCT scans based on light scattering patters in matched pairs of prior and current OCT columns; 4) prior and current OCT column matching via prior and current landmark-based IR image registration; 5) comprehensive automatic progression analysis of AMD atrophy on pairs of consecutive OCT studies; 6) simultaneous visual display of the registered prior and atrophy OCT scan slices and of the atrophy lesions and lesion changes computed from each OCT scan onto the corresponding IR image.

The advantages of our method are: 1) it is fully automatic; 2) it relies on easy-to-generate OCT atrophy columns for the training set; 3) it requires significantly fewer manual annotations that existing methods; 4) it provides a user-friendly viewer, ***OCT-SIM,*** for the simultaneous visualization, delineation, and correction of AMD-related retinal atrophy in OCT scans and IR images, and; 5) it provides clinically relevant measures for the evaluation of dry AMD retinal atrophy progression. Our unique simultaneous column-based OCT scan analysis approach may be applicable for wet retinal atrophy and for various retinal atrophies and pathologies.

The limitations of our studies are that all scans were obtained from a single institution and that the manual atrophy delineation and evaluation was performed by two ophthalmologists from the same institution. Another limitation is the spacing between B-scans: all our OCT scans have dense spacing between consecutive B-scans. However, the spacing may be wider in OCT scans from other sources. In this case, performance may decrease, although a better selection of the various parameter's values may balance this effect. The main limitation of our method, as well as that of all supervised deep-learning methods, is that it requires manually labeled OCT scans by expert ophthalmologists, which may be difficult to obtain. However, with ***OCT-SIM***, we have shown that expert level performance can be achieved with a relatively modest annotation effort by ophthalmologists.

## 6. Conclusion

Automated accurate and reliable radiomics-based progression analysis of consecutive retinal OCT imaging studies of patients with advanced macular degeneration, atrophy associated with AMD, addresses a current unmet clinical need. It provides a computerized tool that may help to alleviate one of the main problems at the ophthalmic clinics, i.e., the high number of patients and the associated retinal images that have to be reviewed by ophthalmologists in every visit, which is currently performed manually. The reduction of time, the detection of subtle changes, and the reduction of observer bias and intra-observer variability can be of great help to improve everyday ophthalmic examinations. Our method may also provide a useful tool for further research and for clinical applications. These include the quantitative longitudinal evaluation of disease progression, the evaluation of the potential efficacy of new treatments, and the development of disease progression prognosis.

**References**

1. D.S. Friedman, B.J. O'Colmain, B. Munoz, S.C. Tomany, S. McCarty, P.T. de Jong, B. Nemesure, P. Mitchell, J. Kempen. Prevalence of age-related macular degeneration in the USA. Arch. Ophthalmol. 122:564-572, 2004.

2. L. Lim, P. Mitchell, J. Seddon, F. Holz, T. Wong. Age-related macular degeneration. The Lancet 379:1728-38, 2012.

3. C. Bowes Rickman, S. Farsiu, C.A. Toth, M. Klingeborn. Dry age-related macular degeneration: mechanisms, therapeutic targets, and imaging. Investigative Ophthalmology & Visual Science 54(1): ORSF68–80, 2013

4. S.R. Sadda, R. Guymer, F.G. Holz, S. Schmitz-Valckenberg. Consensus definition for atrophy associated with age-related macular degeneration on OCT: classification of atrophy. Ophthalmology 125:53748, 2018.

5. F. Coscas, M. Lupidi, J.F. Boulet, A. Sellam, D. Cabral, R. Serra, C. Francais, E.H. Souied, G. Coscas. Optical coherence tomography angiography in exudative age-related macular degeneration: a predictive model for treatment decisions. British J. Ophth. 103:1342-46, 2019.

6. M.W.M. Wintergerst, T. Schultz, J. Birtel, A.K. Schuster, N. Pfeiffer, S. Schmitz-Valckenberg, F.G. Holz, R.P. Finger. Algorithms for the automated analysis of age-related macular degeneration biomarkers on optical coherence tomography: a systematic review. Transactions on Vision Science Technology 6(4):10, 2017.

7. A. Szeskin, R. Yehuda, O. Shmueli, J. Levy, L. Joskowicz. A column-based deep learning method for the detection and quantification of atrophy associated with AMD in OCT scans. Med Image Analysis 72:102130, 2021.

8. S.J. Chiu, J.A. Izatt, R.V. O'Connell, K.P. Winter, C.A. Toth, S, Farsiu. Validated automatic segmentation of AMD pathology including Drusen and Geographic Atrophy in SD-OCT images. Investigative Ophthalmology & Visual Science 53(1):53–61, 2012.

9. S. Niu, L. de Sisternes, Q. Chen, T. Leng, D.L. Rubin. Automated Geographic Atrophy segmentation for SD-OCT images using a region-based CV model via local similarity. Biomed Optics Express 7(2):581–600, 2016.

10. J. Oliveira, S. Pereira, L. Gonçalves, M. Ferreira, C.A Silva. Multi-surface segmentation of OCT images with AMD using sparse high order potentials. Biomedical Optics Express 8(1):281–297, 2017.

11. T. Kurmann, S. Yu, P. Márquez-Neila, A. Ebneter, M. Zinkernagel, M.R. Munk, R. Sznitman. Expert-level automated biomarker identification in OCT scans. Scientific Reports 9(1):1-9, 2019.

12. C.S. Lee, A.J. Tyring, N.P. Deruyter, Y. Wu, A. Rokem, A.Y. Lee. Deep-learning based automated segmentation of macular edema in optical coherence tomography. Biomed Optics Express 8(7):3440-3448, 2017.

13. Z. Ji, Q. Chen, S. Niu, T. Leng, D.L. Rubin. Beyond retinal layers: a deep voting model for automated geographic atrophy segmentation in SD-OCT images. Translational Vision Science & Technology 7(1):1-15, 2018.

14. F.G. Venhuizen, B. van Ginneken, B. Liefers, M.J. van Grinsven, S., Fauser, C. Hoyng, T. Theelen, C.I. Sánchez. Robust total retina thickness segmentation in Optical Coherence Tomography images using Convolutional Neural Networks. Biomed Optics Express 8(7):3292–3316, 2017.

15. J. De Fauw, J.R. Ledsam, B. Romera-Paredes, S. Nikolov, N. Tomasev, S. Blackwell, G. van der Driessche. Clinically applicable deep learning for diagnosis and referral in retinal disease. Nat. Medicine 24(9):1342, 2018.

16. L. Fang, D. Cunefare, C. Wang, R.H. Guymer, S. Li, S. Farsiu. Automatic segmentation of nine retinal layer boundaries in OCT images of non-exudative AMD patients using deep learning and graph search. Biomedical Optics Express 8(5):2732-2744, 2017.

17. A. Shah, L. Zhou, M.D. Abrámoff, X. Wu. Multiple surface segmentation using convolution neural nets: application to retinal layer segmentation in OCT images. Biomed Optics Express 9(9):4509-4526, 2018.

18. J. Dong, J. Chen, X. Gao, R. Xu, S. Niu. Automated geographic atrophy segmentation with multi-loss for SD-OCT images based on patient independent. Proc. 12th Int. Conf. on Graphics and Image Processing Vol. 11720, p. 1172012). Int. Society for Optics and Photonics, 2021.

19. Y. Derradji, A. Mosinska, S. Apostolopoulos, C. Ciller, S. De Zanet, I. Mantel. Fully-automated atrophy segmentation in dry age-related macular degeneration in OCT. Scientific Reports, 11(1):1-11, 2021.

20. H. Bogunović, A. Montuoro, M. Baratsits, M.G. Karantonis, S.M. Waldstein, F. Schlanitz, U. Schmidt-Erfurth. Machine learning of the progression of intermediate age-related macular degeneration based on OCT imaging. Investigative Ophthalmology & Visual Science 58(6), 2017.

21. I. Banerjee, L. de Sisternes, J. Hallak, T. Leng, A. Osborne, M. Durbin, D.L. Rubin. A deep-learning approach for prognosis of Age-Related Macular Degeneration disease using SD-OCT imaging biomarkers. arXiv preprint arXiv:1902.10700, 2019.

22. U. Schmidt-Erfurth, S.M. Waldstein, S. Klimscha, A. Sadeghipour, X. Hu, B.S. Gerendas, A. Osborne, H. Bogunovic. Prediction of individual disease conversion in early AMD using artificial intelligence. Investigative Ophthalmology & Visual Science 59:3199–3208, 2018.

23. A. Birenbaum, H. Greenspan. Multi-view longitudinal CNN for multiple sclerosis lesion segmentation. Engineering Applications of Artificial Intelligence 65:111-8, 2017.

24. L. Vandewinckele, S. Willems, D. Robben, J. Van Der Veen, W. Crijns, S. Nuyts, F. Maes, F. Segmentation of head-and-neck organs-at-risk in longitudinal CT scans combining deformable registrations and convolutional neural networks. Computer Methods in Biomechanics and Biomedical Engineering 8(5):519-28, 2020.

25. C. Stewart, C. Tsai, B. Roysam,. The dual-bootstrap iterative closest point algorithm with application to retinal image registration. IEEE Trans. Med. Imag. 22(11): 1379-1394, 2003.

26. A. Frangi, W. Niessen, K. Vincken, M. Viergever, 1998. Multiscale vessel enhancement filtering. Lecture Notes in Computer Science, LNCS Vol 1496,130-137, Springer-Verlag. Germany.

27. E. Rublee, V. Rabaud, K. Konolige, G. Bradski. ORB: an efficient alternative to SIFT or SURF. Proc. IEEE Int. Conf. Computer Vision, pp. 2564-2571, 2011.

28. M. Pfau, L. von der Emde, L. de Sisternes, J.A. Hallak, T. Leng, S. Schmitz-Valckenberg, F.G. Holz, M. Fleckenstein, D.L. Rubin. Progression of photoreceptor degeneration in Geographic Atrophy secondary to AMD. JAMA Ophthalmol. 138(10):1026-1034, 2020.

29. C.J. van Rijsbergen, 1979. Information Retrieval. London: Butterworths.

30. M. Treder, J.L. Lauermann, N. Eter. Deep learning-based detection and classification of geographic atrophy using a deep convolutional neural network classifier. Graefe's Archive for Clinical and Experimental Ophtalmology 256(11):2053-2060, 2018.

**Figure legends, tables, and figures**

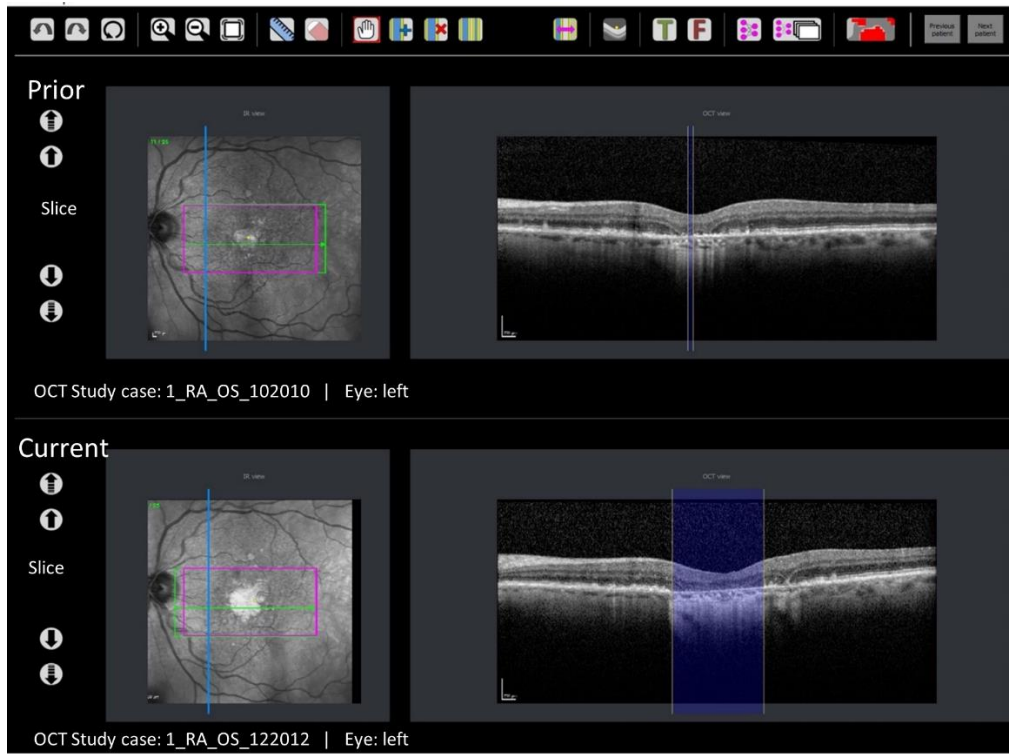

**Fig. 1.** *OCT-SIM* viewer screen showing registered prior (top) and current (bottom) OCT studies of a patient with dry AMD taken 26 months apart. Each OCT study consists of an infrared (IR) image (left) and an OCT scan (OCT slice, right). Overlaid on the IR images are the OCT scan field of view (magenta rectangle) and the location of OCT slice on display (green line). Overlaid on the OCT slices are cRORA segments (blue) whose boundaries are the leftmost and rightmost columns (yellow lines). The increased degeneration of the RPE layer inside the atrophy segments manifest by the light scattering pattern. The increase in the atrophy segment width from the prior to the current OCT scan matched slices indicates cRORA progression.

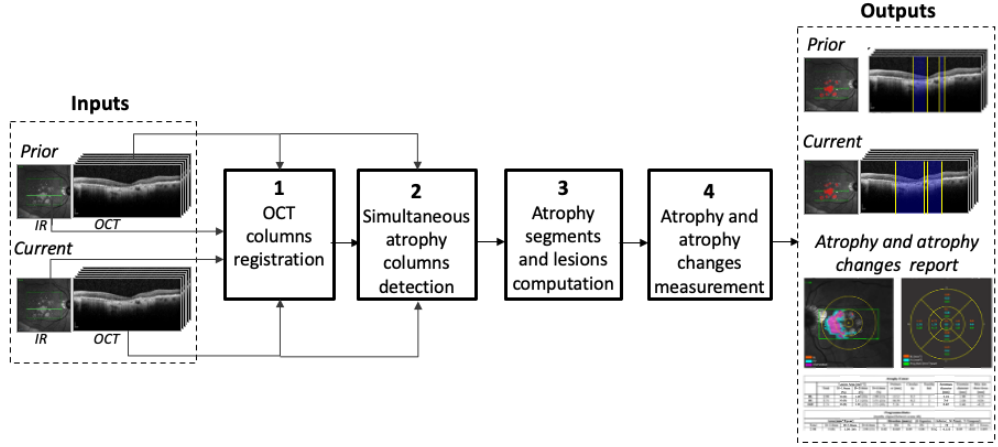

**Fig. 2.** Flowchart of the retinal atrophy changes analysis pipeline in pairs of OCT studies. Inputs: prior and current OCT studies (IR image and OCT scan); outputs: matching OCT prior and current OCT slices, atrophy segments in the prior and current OCT scans (blue segments delimited by yellow lines), the atrophy lesions in the corresponding IR images (red), and the atrophy lesions and an atrophy changes report. The steps are: 1) registration of the prior and the current OCT columns via their IR images; 2) simultaneous detection of the prior and current atrophy columns with a simultaneous column-based model; 3) computation of the prior and current atrophy segments in the OCT scans and the atrophy lesions in the IR images; 4) computation of the atrophy and atrophy changes report.

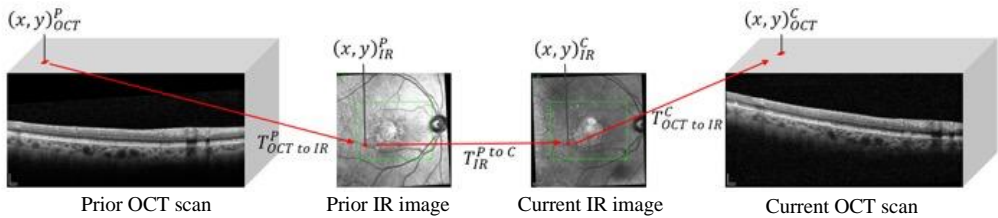

**Fig. 3.** OCT column registration transformations (red arrows): an OCT column in the prior OCT scan (left) with coordinates $(x, y)_{OCT}^P$ is related to a pixel $(x, y)_{IR}^P$ in the prior IR image by transformation $T_{OCT\ to\ IR}^P$. An OCT column in the current OCT

scan (right) with coordinates $(x,y)^C_{OCT}$ is related to a pixel $(x,y)^C_{IR}$ in the current IR image by transformation $T^C_{OCT\ to\ IR}$. Transformation $T^{P\ to\ C}_{IR}$ matches pixel locations in the prior and current images. The closest matching prior and current OCT slices are obtained from the prior and current $x$ coordinates.

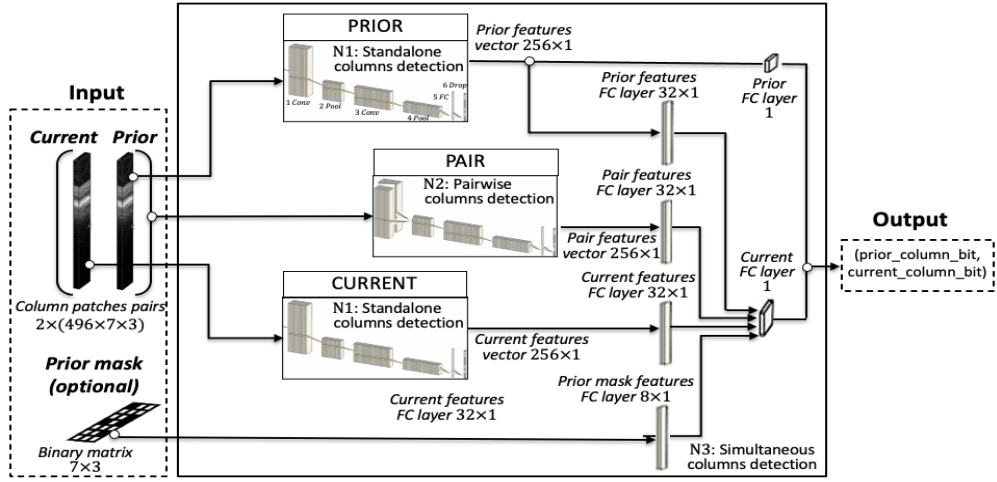

**Fig. 4**: Architecture of the simultaneous multi-channel column-based CNN network N3. Input: a pair of registered prior and current 3D OCT column patches (left. top), and optionally, the prior atrophy segments matrix (left bottom). Output: a pair of bits indicating the absence/presence (0/1) of atrophy in the input prior and current OCT columns (right, center). The network consists of two identical standalone column feature extraction (FE) networks N1 (PRIOR, CURRENT) and a pairwise column FE network N2 (PAIR). The FE networks output features vectors (prior, current and pairs) of size 256×1; they are connected with fully connected (FC) layers whose outputs are vectors of size 32×1. Optionally, the prior mask is connected by an FC layer with a vector size of 32×1 followed by a FC layer with a vector size of 8×1. The prior column bit is computed by a single-channel FC layer; the current column classification bit is computed with a flattened four-channel FC

layer. The standalone column FE network N1 consists of two sequences of a convolutional and a max-pooling layers followed by an FC and a drop-out layer. The pairwise column FE network N2 is identical to N1 except for its first convolution layer, which inputs pairs of 3D column patches. The numbers indicate the kernel sizes.

|  | PRIOR Ground Truth | CURRENT Ground Truth | CURRENT ||||
| --- | --- | --- | --- | --- | --- | --- |
|  |  |  | Standalone | M3_M2 | M3_M1 | M3 |
| CASE 1 | | | | | | |
| | | | 0.69 | 0.77 | 0.77 | 0.82 |
| CASE 2 | | | | | | |
| | | | 0.75 | 0.78 | 0.79 | 0.79 |

**Fig. 5**. Examples of the results of two OCT studies (Case 1, Case 2) from the test set of the standalone and the three simultaneous column-based models M3_M2, M3_M2, and M3 for cRORA. The first two columns show for each case, the original (top) IR image cropped to the OCT scan FOV with the ground truth atrophy lesion segmentation masks (bottom) prior (red) and current (blue) overlays. The next four columns show the current IR images with the atrophy lesions segmentation masks (blue) computed with the four classifiers and their $F_1$ scores (lower right R images).

| MEASUREMENT DIFFERENCES | | Change in focality index | Area progression rate [mm$^2$/year] | Mean outward progression rate [mm/year] | Mean inward progression rate [mm/year] | Mean Feret progression rate [mm/year] |
|---|---|---|---|---|---|---|
| Inter-observer | **RMSE** | **1.76** | **0.74** | **0.44** | **0.11** | **0.64** |
| | std | 1.75 | 0.73 | 0.41 | 0.10 | 0.57 |
| | min | -2.00 | -0.94 | -0.37 | 0.00 | -0.50 |
| | max | 4.00 | 1.74 | 1.00 | 0.32 | 1.54 |
| *M3* without prior mask | **RMSE** | **1.29** | **0.90** | **0.18** | 0.06 | **0.29** |
| | std | 0.82 | 0.64 | 0.16 | 0.06 | 0.27 |
| | min | 0.00 | -1.25 | -0.11 | -0.13 | -0.16 |
| | max | 2.00 | 0.61 | 0.36 | 0.06 | 0.63 |

**Table 1.** Differences for the five most significant clinical measurements. Listed for each measurements is the RSME, std, minimum and maximum differences computed from the atrophy lesion segmentations. The measurement differences are computed for the two manual annotations (inter-observer differences) and for those computed with model M3 without a prior mask vs. those computed from the ground truth annotations.

| Models With/ Without prior mask | Measure | ATROPHY SEGMENTATION | | | | | | ATROPHY DETECTION | | | |
|---|---|---|---|---|---|---|---|---|---|---|---|
| | | | | | | | | SEGMENT | | LESION | |
| | | Ground truth *columns* | Computed *columns* | Burden Diff % mean (std) | $F_1$ score mean (std) | ASSD mean (std) | SHD mean (std) | Precision mean (std) | Recall mean (std) | Precision mean (std) | Recall mean (std) |
| *M1* Standalone | Total | 40,587 | 56,254 | +10.0 (10.7) | 0.72 (0.07) | 15.4 (20.1) | 84.9 (97.1) | 0.70 (0.16) | 0.94 (0.06) | 0.66 (0.17) | 0.89 (0.25) |
| | Mean (std) | 6,764 (2,371) | 9,375 (2,968) | | | | | | | | |
| *M3_M2_P* with prior mask | Total | 40,587 | 53,919 | +8.3 (6.5) | 0.75 (0.06) | 4.5 (3.7) | 4 6.6 (28.5) | 0.90 (0.06) | 0.89 (0.07) | 0.84 (0.17) | 0.89 (0.16) |
| | Mean (std) | 6,764 (2,371) | 8,986 (2,648) | | | | | | | | |
| | Total | 40,587 | 52,265 | +7.3 | 0.74 | 4.6 | 44.1 | 0.88 | 0.89 | 0.57 | 0.78 |
| **M3_M2** without prior mask | Mean (std) | 6,764 (2,371) | 8,710 (2,503) | (7.2) | (0.06) | (4.2) | (32.1) | (0.08) | (0.06) | (0.25) | (0.31) |
| **M3_M1_P** with prior mask | Total Mean (std) | 40,587 6,764 (2371) | 57,877 9,646 (2,371) | +10.5 (6.6) | 0.75 (0.06) | 4.7 (3.8) | 41.4 (30.5) | 0.90 (0.06) | 0.92 (0.06) | 0.77 (0.17) | 0.94 (0.12) |
| **M3_M1** without prior mask | Total Mean (std) | 40,587 6,764 (2,371) | 65,107 10,851 (4262) | +14.5 (8.8) | 0.74 (0.08) | 10.7 (12.6) | 64.7 (67.9) | 0.85 (0.12) | 0.96 (0.03) | 0.63 (0.20) | 0.94 (0.12) |
| **M3_P** with prior mask | Total Mean (std) | 40,587 6764 (2371) | 172,077 14,340 (6,574) | +8.2 (7.8) | 0.77 (0.07) | 4.5 (4.1) | 45.7 (25.7) | 0.91 (0.06) | 0.89 (0.06) | 0.84 (0.17) | 0.83 (0.25) |
| **M3** without prior mask | Total | 40,587 | 62,249 | +13.0 (6.7) | 0.74 (0.08) | 4.5 (3.5) | 33.9 (31.1) | 0.90 (0.09) | 0.95 (0.05) | 0.74 (0.18) | 0.94 (0.12) |

**Table 2.** Results of the standalone (M1) and the simultaneous column-based models for cRORA with and without prior mask: M3_M2_P and M3_M2 (no pairwise column FE network N2), M3_M1_P and M3_M1 (no prior and current column FE networks N1), and M3_P and M3 (full simultaneous networks). Columns 2-8 list the atrophy segmentation: the number of columns (total, mean, std) in the ground truth and in the computed segmentation, their difference in terms of atrophy burden percentage (the ration between the total atrophy area and the total FOV area), and their mean (std) similarity: F1 score, ASSD, and SHD in mm. Columns 9-12 list the atrophy detection precision and recall for atrophy segments and atrophy lesions.